\documentclass[twocolumn]{revtex4}%
\usepackage{amssymb}
\usepackage{amsmath}
\usepackage{amsfonts}
\usepackage{graphicx}
\usepackage{dcolumn}
\usepackage{epstopdf}

\begin{document}

\title{Nuclear physics in the cosmos}

\author{Carlos A. Bertulani}
\affiliation{ Department of Physics and Astronomy, Texas A\&M University-Commerce, Commerce,
TX 75429, USA\footnote{Email: carlos\_bertulani@tamu-commerce.edu\\XXXIV edition of the Brazilian Workshop on Nuclear Physics, 5-10 June 2011, Foz de Igua\c{c}u, Parana state, Brazil}}

\begin{abstract}
We observe photons and neutrinos from stars. Based on these observations, complemented by measurements of cosmic rays energies and composition, we have been able to constrain several models for the Big Bang and for stellar evolution.  But that is not enough. We also need to help this effort with laboratory experiments. We are still far from being able to reproduce stellar environments in a terrestrial  laboratory. But in many cases we can obtain accurate nuclear reaction rates needed for modeling primordial
nucleosynthesis and hydrostatic burning in stars. The relevant
reactions are difficult to measure directly in the
laboratory at the small astrophysical energies. In recent years
indirect reaction methods have been developed and applied to extract
low-energy astrophysical S-factors. These methods require a
combination of new experimental techniques and theoretical efforts,
which are the subject of this short review.

\end{abstract}
		
\maketitle

\section{Astrophysics: what we can and what we can't do}

\subsection{Hot plasmas on Earth}

Evidently, we cannot  reproduce in the laboratory conditions existing during the Big Bang and during stellar evolution. But efforts to reproduce such conditions on a limited scale on Earth are underway. A good example are experiments being carried out at the National Ignition Facility (NIF) in Livermore. In this facility
the intense energy of 192 giant laser beams is focused on a small spherical pellet containing a few milligrams of fusion fuel, typically a mix of deuterium and tritium. The energy  heats the surface of the pellet into a plasma,  exploding off its surface, driving the remaining portion of the target  inwards, and compressing it into a  high density. A shock wave  travels towards the center of the compressed fuel from all sides, further heating and compressing it so that fusion reactions will occur and release energy, creating temperatures and pressures similar to those that exist only in the cores of stars and giant planets and inside nuclear weapons \cite{Li04}.

Another example is ITER, a large-scale international laboratory located in France that aims to demonstrate that it is possible to produce commercial energy from fusion. ITER is based on the ``tokamak" concept of magnetic confinement, in which the plasma is contained in a doughnut-shaped vacuum vessel. A mixture of deuterium and tritium is heated to temperatures of 150 million $^\circ$C, forming a hot plasma. Strong magnetic fields are used to keep the plasma away from the walls. From 50 MW of input power, the ITER machine is designed to produce 500 MW of fusion power. ITER runs on a predicted 15 billion euros building cost, whereas NIF already costs roughly US\$ 5 billions. So, these are not cheap machines at all. It is very hard to reproduce conditions within stars. And the prospects of generating energy for commercial use with similar projects in the future are still uncertain. 
As for helping us understanding features of the Big Bang and of stellar evolution, ITER will not be able to tell us much. It will mainly access questions on atomic and material science associated with confining a plasma at huge temperatures within a vessel and the interactions of the plasma with the walls of the vessel. While it is undeniable that this experiment will fill a knowledge gap needed for further developments in science, it will not answer crucial questions of relevance for astrophysics \cite{Gar11}. Maybe ITER-2 will, if we can afford it. 

To avoid sounding too negative, I mention that NIF has a good plan to provide results on atomic and nuclear physics for stellar evolution. I give a couple of examples. In the theoretical modeling of stellar evolution one relies strongly on calculations of radiation propagation through hot stellar plasmas. The coefficient entering the radiation propagation equation is called the ``opacity". It accounts for the interaction of photons with atoms and effects such as excitation and ionization of ground-state, excited, or ionized atomic species present in the medium. For many years we have relied on a huge effort to calculate all of the atomic physics needed for stellar evolution codes in the form of opacity tables \cite{Ig96}. Stellar modelers have not questioned much the reliability of such tables, as one simply can't do better than that. But it would be a great knowledge improvement if we could effectively ``measure" opacity in the laboratory.  The NIF X-ray opacity platform will enable detailed studies of the radiative properties of hot dense matter  over a photon energy range of 200 - 10,000 eV,  also important  in astrophysics \cite{GHB08}. It will allow benchmarking opacities used in the standard solar model and in stellar equilibrium codes (relevant to exoplanet habitability assessment) and absorption/emission spectroscopy of photoionized plasmas scaled to black hole and neutron star accretion-disk conditions. The development of pulsed power and high power lasers opens a brand new perspective for the study of opacities in several dense plasmas including modeling of the atmospheres of very cool white dwarf stars \cite{Win11}.  

\section{Nuclear reactions}

Stars are powered by nuclear reactions at very low energies and, in many situations, at very high densities. Usually, one needs to know what happens during binary encounters between nuclei (a counter-example is the celebrated triple-$\alpha$ reaction). The effects of the environment electrons are still a disputed research topic. But the main problem here is really to know the reaction rates at the energies required for stellar modeling. For example, in our Sun the reaction $^{7}$Be$\left(  {\rm p},\gamma\right)
^{8}$B plays a major role for the production of high energy
neutrinos from the $\beta$-decay of $^{8}$B. These neutrinos come
directly from the center of the Sun and are ideal probes of the
sun's structure. John Bahcall frequently said that this was
the most important reaction in nuclear astrophysics \cite{John}. Our
knowledge about this reaction has improved considerably due to new
radioactive beam facilities. Another example, the reaction $^{12}$C$\left(
\alpha,\gamma\right)  ^{16}$O, is extremely relevant for the fate of
massive stars. It determines if the remnant of a supernova explosion
becomes a black-hole or a neutron star. These two reactions are just
two examples of a large number of reactions which are not yet known
with the required accuracy needed in astrophysics.

NIF has reported the first cross section and spectral measurements  of the T(t,2n)$^4$He reaction that is an important mirror reaction to the $^3$He($^3$He,2p)$^4$He reaction (which is part of the proton-proton chain in hydrogen burning stars). These direct measurements, which were conducted at energies inaccessible by conventional accelerator-based techniques, are not affected by electron screening. Measurements of the differential cross section for the elastic n-$^3$H and n-$^2$H scattering at 14.1 MeV have also been published \cite{Fre11}.  The accurate determination of this reaction rate is essential for understanding how the fuel is assembled in an implosion, and for  the demonstration of thermonuclear ignition and net energy gain at NIF.  It also opens the door for planning the use of NIF and other laser powered facilities to obtain information on nuclear reaction rates at the energies occurring in stars. 

The extremely low cross sections for reactions induced by charged particles and the inherent difficulty to obtaining reaction cross sections induced by low energy neutrons leads to enormous hurdles to develop reliable stellar evolution models and computer codes.  Chains of low energy nuclear reactions lead to complicated
phenomena such as nucleosynthesis, supernovae explosions, and energy
production in stars. An example is that approximately half of all stable nuclei observed in nature in the
heavy element region, $A>60$, are produced during the ``r--process". The exact site of the r--process is 
not known, but one believes that it  occurs in environments with large neutron densities 
leading to neutron capture times much smaller than the beta-decay
half--lives, $\;\tau _{\mathrm{n}}\ll\tau_{\beta}$, of the nuclei involved. The most
neutron--rich isotopes along the r--process path have lifetimes of
less than one second; typically 10$^{-2}$ to 10$^{-1}$\thinspace s.
Cross sections for most of the nuclei involved are hard to measure
experimentally. Sometimes, theoretical calculations of the capture
cross sections and of the beta--decay half--lives are the only
source of input for r--process modeling.

Nucleosynthesis in stars is also complicated by the
presence of electrons. They screen the nuclear charges, therefore increasing
the fusion probability by reducing the Coulomb repulsion. Evidently, the
fusion cross sections measured in the laboratory have to be corrected by the
electron screening when used in a stellar model. This is a purely
theoretical problem as one can not exactly reproduce the conditions at stellar interiors in the
laboratory. At least for now.

A simpler screening mechanism occurs in laboratory experiments due to the
bound atomic electrons in the nuclear targets. This case has been studied in
great detail experimentally, as one can control different charge states of
the projectile+target system in the laboratory
\cite{Ass87,Rol95,Rol01,Kas02,Cze06}. The experimental findings disagree
systematically by a factor of two or more with theory. This is surprising as the
theory for atomic screening in the laboratory relies on our basic knowledge of
atomic physics. At very low energies one can use the simple adiabatic model in
which the atomic electrons rapidly adjust their orbits to the relative motion
between the nuclei prior to the fusion process. Energy conservation requires
that the larger electronic binding (due to a larger charge of the combined
system) leads to an increase of the relative motion between the nuclei, thus
increasing the fusion cross section. As a matter of fact, this enhancement has
been observed experimentally. The measured values are however not compatible
with the adiabatic estimate \cite{Ass87,Rol95,Rol01,Kas02,Cze06}. Dynamical
calculations have been performed, but they obviously cannot explain the
discrepancy as they include atomic excitations and ionizations which reduce
the energy available for fusion. Other small effects, like vacuum
polarization, atomic and nuclear polarizabilities, relativistic effects, etc.,
have also been considered \cite{BBH97}. But the discrepancy between experiment
and theory remains \cite{BBH97,Rol01}.

A possible solution of the laboratory screening problem was proposed
\cite{LSBR96,BFMH96}.
Experimentalists often use the extrapolation of stopping power tables \cite{AZ77} to obtain the average value of
the projectile energy due to stopping in the target material. The
stopping is due to ionization, electron-exchange, and other atomic
mechanisms. However, the extrapolation is challenged by theoretical
calculations which predict a lower stopping. Smaller stopping was
indeed verified experimentally \cite{Rol01}. At very low energies,
it is thought that the stopping mechanism is mainly due to electron
exchange between projectile and target. This has been studied in
Ref. \cite{BD00} in the simplest possible situation: proton+hydrogen
collisions. The calculated stopping power was added to the nuclear
stopping power mechanism, i.e. to the energy loss by the Coulomb
repulsion between the nuclei. The obtained stopping power is
proportional to $v^{\alpha}$, where $v$ is the projectile velocity
and $\alpha=1.35$. The extrapolations from stopping power
tables predict a smaller value of $\alpha$. Although this result
seems to indicate the stopping mechanism as a possible reason for
the laboratory screening problem, the theoretical calculations tend
to disagree on the power of $v$ at low energy collisions
\cite{GS91}.

Another calculation of the stopping power in atomic He$^{+}+$He
collisions using the two-center molecular orbital basis was reported in
Ref. \cite{Ber04}.  The agreement with the data from Ref.
\cite{GS91} at low energies is excellent. The agreement with the
data disappears if nuclear recoil is included. In fact, the
unexpected ``disappearance" of the nuclear recoil was also observed
in Ref. \cite{Form}. This seems to violate a basic principle of
nature, as the nuclear recoil is due to Coulomb repulsion between
projectile and target atoms \cite{AZ77}. After several attempts, sometimes with elaborate theoretical models,
little theory activity in this field has been reported.  Some models have been praised as solving the stellar screening 
problem (see, e.g. \cite{BS97}). I believe that this praise is more due to the use of quantum-field theoretical tools, 
which tends to impress low-energy  experimentalists and theorists who know little about those theoretical techniques. 
The fact is that the present situation on screening of nuclear reactions is confusing. Either experimentalists are publishing wrong analysis, or all aspects of theory 
might not have been considered yet \cite{RMP11}.   

I have discussed above a few ongoing efforts in nuclear astrophysics, some expensive investments to get solutions, as well as some problems that might not even be possible to study on Earth.  Actually, much of the knowledge required for understanding  the physics of the Big Bang and of stellar evolution  can be accessed by means of indirect methods in nuclear physics. The goal of these theoretical methods, and laboratories that use them, is to take a detour following a much harder work of putting pieces together from several,  sometimes seemingly unrelated, experiments. In this article I will review some of these methods and what has been and can be accomplished   with them.

\section{Understanding fusion cross sections}

\subsection{Fusion}

 All approaches to understand fusion reactions involve two prongs: a) Calculate an ion-ion (usually one-dimensional) phenomenological potential (Wood-Saxon, proximity, folding, Bass, etc.) using frozen densities, or microscopic, macroscopic-microscopic methods using collective variables (CHF, ATDHF, empirical methods), and b) employ quantum mechanical tunneling methods for the reduced one-body problem (WKB, IWBC),  incorporating quantum mechanical processes by hand, including neutron transfer and excitations of the entrance channel nuclei (CC). Only for very light ions, involving nuclei lighter than oxygen it is possible to devise more microscopic methods, based on binary nucleon-nucleon interactions, to obtain the fusion reaction cross sections of interest for nuclear astrophysics \cite{QN08}.
 
Fusion cross sections can be calculated from the equation
\begin{equation}
\sigma_F(E)=\pi {\lambda}^2 \sum_\ell (2\ell +1) P_\ell (E) , \label{eqnf}
\end{equation}
where $E$ is the center of mass energy, $\lambda=\sqrt{\hbar^2/2mE}$ is the reduced wavelength and $\ell=0,1,2,\cdots$.
The cross section is proportional to $\pi\lambda^2$, the area of the quantum wave. Each part of the wave corresponds to different impact parameters having different probabilities for fusion. As the impact parameter increases, so does the angular momentum, hence the reason for the $2\ell+1$ term. $P_\ell(E)$ is the probability that  fusion occurs at a given impact parameter, or angular momentum. The barrier penetration method (BPM) assumes that fusion occurs when the particle (with mass $m$) penetrates the Coulomb barrier and $P_\ell$ is calculated in a one-dimensional potential model, e.g. by using the WKB approximation or alike. 
From $\sigma_\ell =\pi {\lambda}^2 (2\ell+1) P_\ell$ one can calculate the average value of $\ell$ from $\langle \ell(E)\rangle = \sum_\ell \ell \sigma_\ell/\sum \sigma_\ell$ and many other relevant quantities.  Sometimes, for a better visualization, or for extrapolation to low energies, one uses the concept of {\it astrophysical S-factor}, redefining the cross section as   
\begin{equation}
\sigma_F(E)={1\over E} S(E)\exp\left[ -2\pi \eta(E)\right], \label{SE}
\end{equation} 
where $\eta(E)=Z_1Z_2e^2/\hbar v$, with $v$ being the relative velocity. The exponential function is an approximation to $P_0(E)$ for a square-well nuclear potential plus Coulomb potential, whereas the factor $1/E$ is proportional to the area appearing in Eq. \ref{eqnf}.   

\begin{figure}
\begin{center}
\resizebox{0.65\columnwidth}{!}{%
\includegraphics{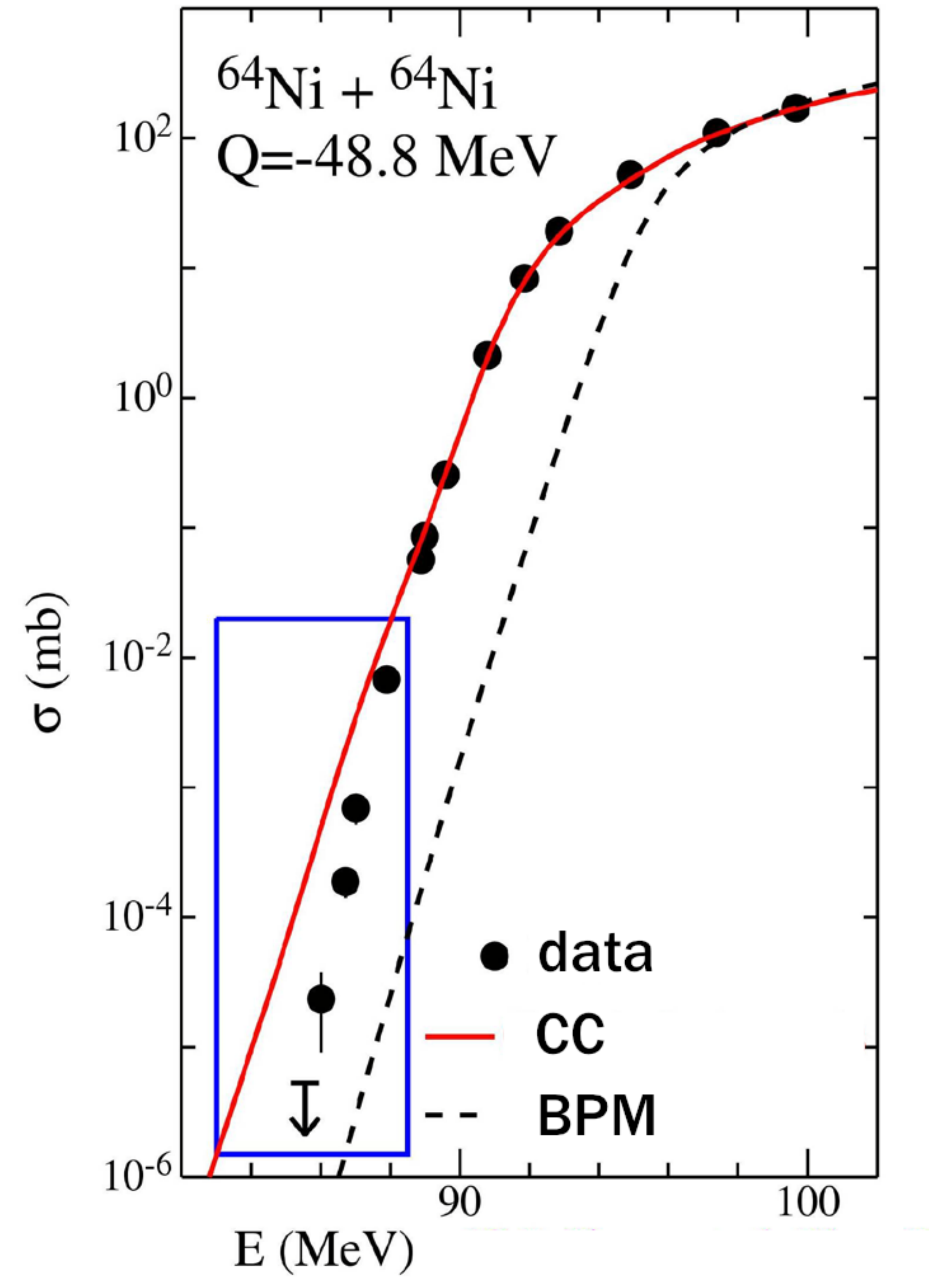}}
\resizebox{1\columnwidth}{!}{\includegraphics{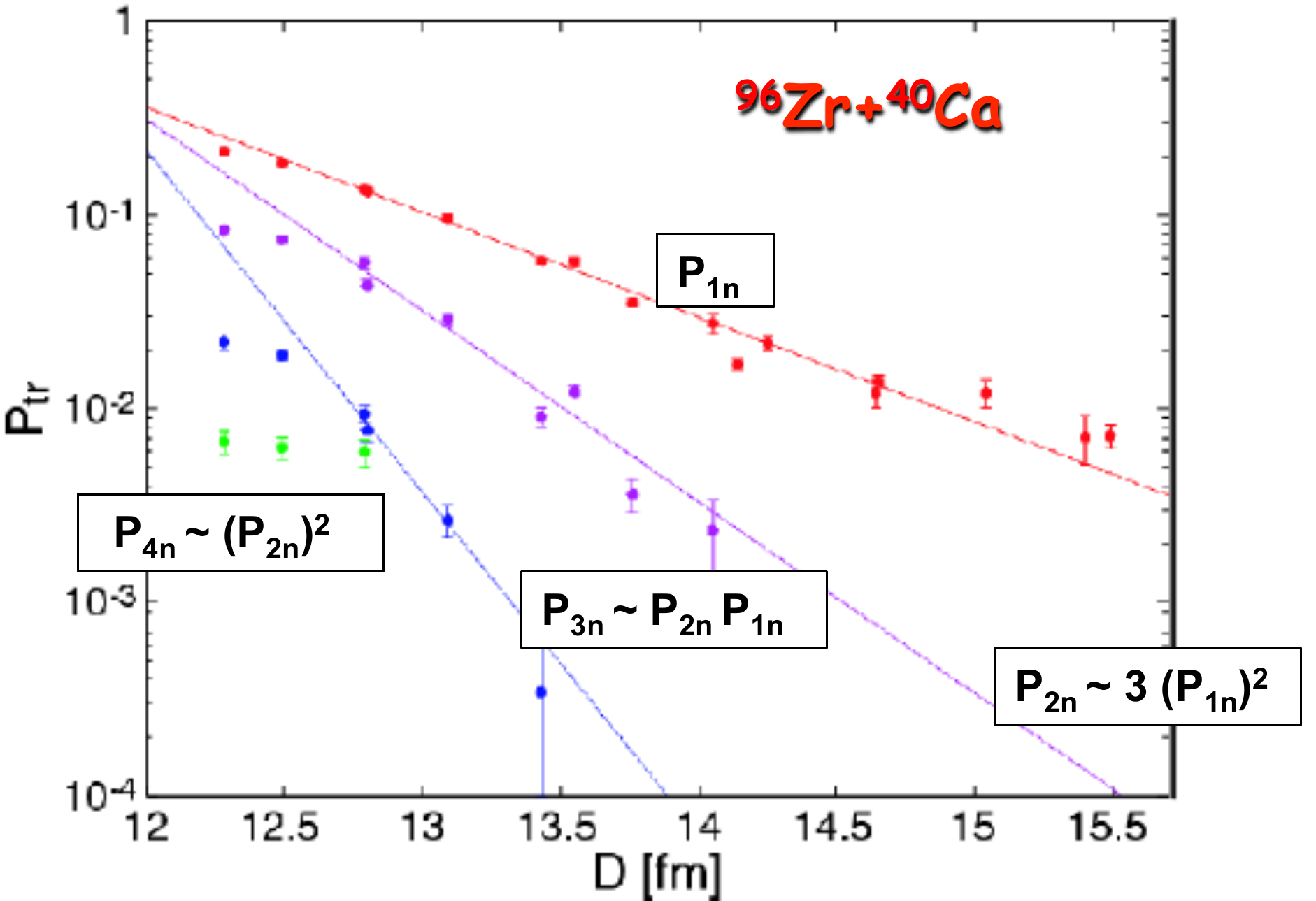}}
\end{center}
\caption{{\it Top} - Fusion cross section of $^{64}$Ni+$^{64}$Ni as a function of the center of mass energy \cite{Ji04}. The dashed (solid) curve is a BPM (coupled-channel) calculation. {\it Bottom} - Transfer probabilities for multineutron transfer in $^{96}$Zr+$^{40}$Ca \cite{Corradi}.}
\label{fig:1}       
\end{figure}

In order to use Eq. \ref{eqnf} one needs the nucleus-nucleus potential. This is a badly known beast. It includes the effects of  non-fusion channels, which might be hardly known. As it cannot be calculated from first principles,  one adds an imaginary part to the real potential and it becomes much more than a beast; something really abnoxious.  Some have tried to tame this thing from first principles. But, except for few heroic attempts, we seem to have given up. We just  fit whatever we can fit and we get whatever parameters of a potential function  we can. Then we simply call it the ``optical potential".

\subsection{Many reaction channels}

The situation is worse, as Eq. \ref{eqnf} does not work in most situations. A good example is shown in figure \ref{fig:1} (top), taken from Ref. \cite{Ji04}. Only by including coupling to other channels, the fusion cross sections can be reproduced.  In  coupled channels schemes  one expands the total wavefunction for the system as
\begin{equation}
\Psi=\sum_{i,k} a_i(\alpha, q_k)\phi(\alpha, q_k),
\end{equation} 
where $\phi$ form the channel  basis, $\alpha$ is a dynamical variable (e.g., the distance between the nuclei), and $q_k$ are intrinsic coordinates. Inserting this expansion in the Schr\"odinger equation yields  a set of CC equations in the form
\begin{equation}
{da_k\over d\alpha}=\sum_j a_j  \ \langle \phi_k \left| U \right| \phi_j \rangle \ e^{E_\alpha \alpha/\hbar}, \label{cc}
\end{equation}
where $U$ is whatever potential couples the channels $k$ and $j$ and $E_\alpha=E_\alpha^{(k)}-E_\alpha^{(j)}$ is some sort of transition energy, or transition momentum. In the presence of continuum states, continuum-continuum coupling (relevant for breakup channels) can be included by discretizing the continuum. This goes by the name of Continuum Discretized Coupled-Channels (CDCC) calculations. There are several variations of CC equations, e.g., a set of differential equations for the wavefunctions, instead of using basis amplitudes. Coupled channels calculations with a large number of channels in continuum couplings, is one of the least controllable calculations. Anything can happen because of the phases of matrix elements: the couplings can add destructively or constructively, depending on the system and on the nuclear model. Such suppressions or enhancements are difficult to understand. 

\subsection{Radiative capture}

For reactions involving light nuclei, only a few channels are of relevance. In his case, a real potential  is enough for the treatment of fusion. For example,  radiative capture cross sections of the type $n+x\rightarrow a+\gamma$
and $\pi L$ ($\pi=E,(M)=$electric (magnetic) L-pole) transitions
can be calculated from (see, e.g., \cite{HBG10})
\begin{equation}
\sigma_{EL,J_{b}}^{\rm d.c.}   = const. \times \left\vert \left\langle l_cj_c\left\Vert
\mathcal{O}_{\pi L}\right\Vert l_{b}j_{b} \right\rangle
\right\vert^{2},
\label{respf}%
\end{equation}
where $\mathcal{O}_{\pi L}$ is an EM operator, and $\left\langle
l_cj_c\left\Vert \mathcal{O}_{\pi L}\right\Vert l_{b}j_{b}
\right\rangle$ is a multipole matrix element involving bound ($b$) and continuum ($c$) wavefunctons. For 
electric multipole transitions ($ \mathcal{O}_{\pi L}= r^LY_{LM}$),
\begin{equation}
\left\langle l_cj_c\left\Vert
\mathcal{O}_{EL}\right\Vert l_{b} j_{b}\right\rangle
=const. \times \int_{0}^{\infty}dr \
r^{L}u_{b}(r)u_{c}(r)
,\label{lol0}%
\end{equation}
where $u_i$ are radial wavefunctions.
The total direct capture cross section is obtained by adding all
multipolarities and final spins of the bound state ($E\equiv E_{nx}$),
\begin{equation}
\sigma^{{\rm d.c.}} (E)=\sum_{L,J_{b}} (SF)_{J_{b}}\ \sigma^{{\rm d.c.}%
}_{L,J_{b}}(E) \ , \label{SFS}%
\end{equation}
where $(SF)_{J_{b}}$ are spectroscopic factors.

\subsection{Asymptotic normalization coefficients}

In a microscopic approach,
instead of single-particle wavefunctions one often makes use of overlap
integrals, $I_{b}(r)$, and a many-body wavefunction for the relative
motion, $u_{c} (r)$. Both $I_{b}(r)$ and $u
_{c}(r)$ might be very complicated to calculate, depending on how
elaborated the microscopic model is. The variable $r$ is the relative
coordinate between the nucleon and the nucleus $x$, with all the intrinsic
coordinates of the nucleons in $x$ being integrated out. The direct capture
cross sections are obtained from the calculation of $\sigma_{L,J_{b}%
}^{{\rm d.c.}} \propto|\left<  I_{b}(r)||r^{L}Y_{L}|| \Psi_{c}(r)\right>
|^{2}$.

The imprints of many-body effects will eventually disappear at large distances
between the nucleon and the nucleus. One thus expects that the overlap
function asymptotically matches ($r\rightarrow\infty$),
\begin{eqnarray}
I_{b}(r)    &=&C_{1} \frac{W_{-\eta,l_{b}+1/2}(2\kappa r)}{r}
\   {\rm for \ protons},  \nonumber \\
I_{b}(r)  &=&C_{2} \sqrt{\frac{2\kappa}{r}}K_{l_{b}+1/2}(\kappa r)   \ {\rm for
\ neutrons}, \label{whitt}%
\end{eqnarray}
where the binding energy of the $n+x$ system is related to $\kappa$ by means
of $E_{b}=\hbar^{2}\kappa^{2}/2m_{nx}$, $W_{p,q}$ is the Whittaker function
and $K_{\mu}$ is the modified Bessel function. In      Eq. \ref{whitt}, $C_{i}$ is
the asymptotic normalization coefficient (ANC).

\begin{figure}
\includegraphics[
height=2 in]{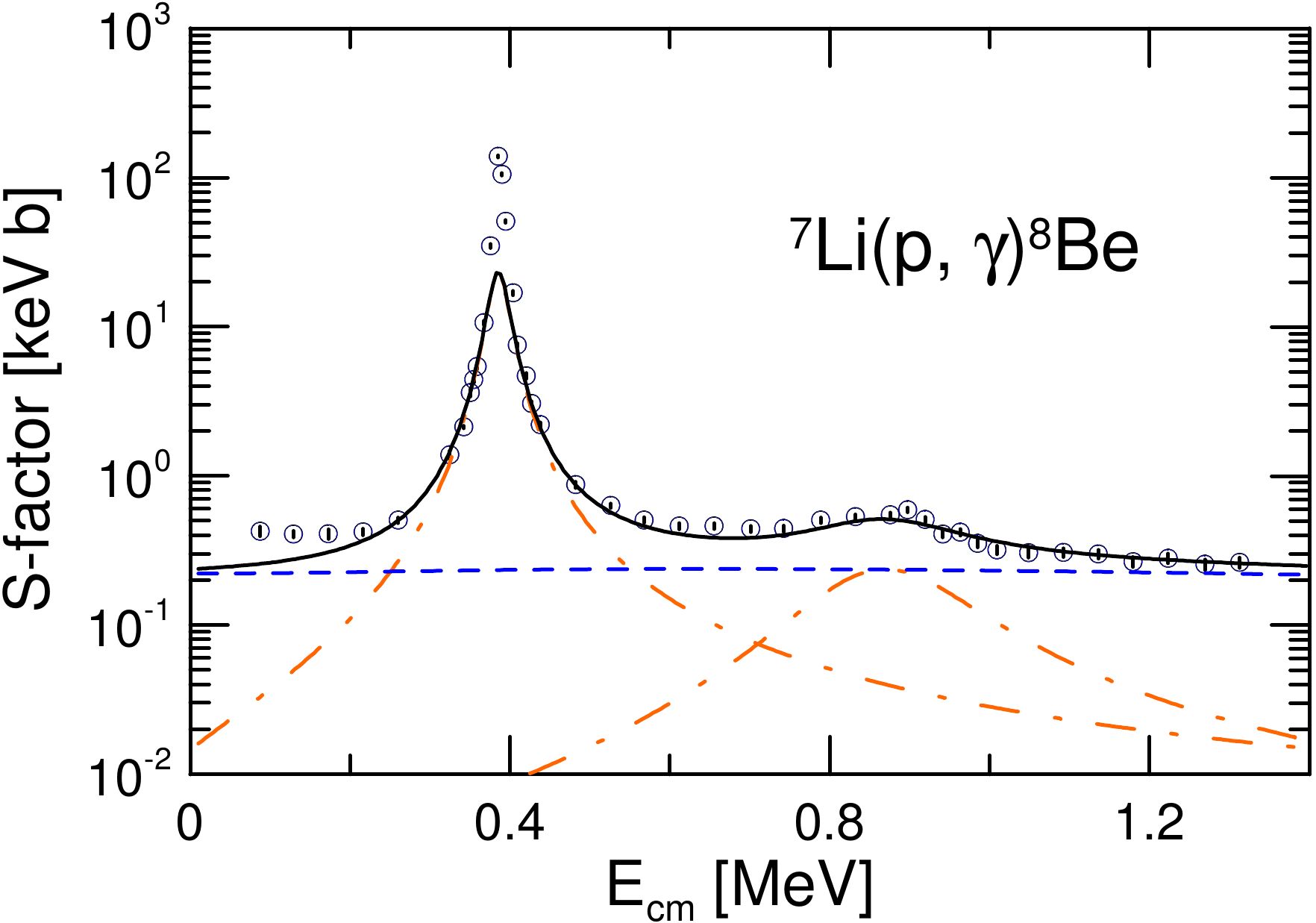}\ \ \ \ \ \ \ \ \ \ \
\includegraphics[
height=2 in]{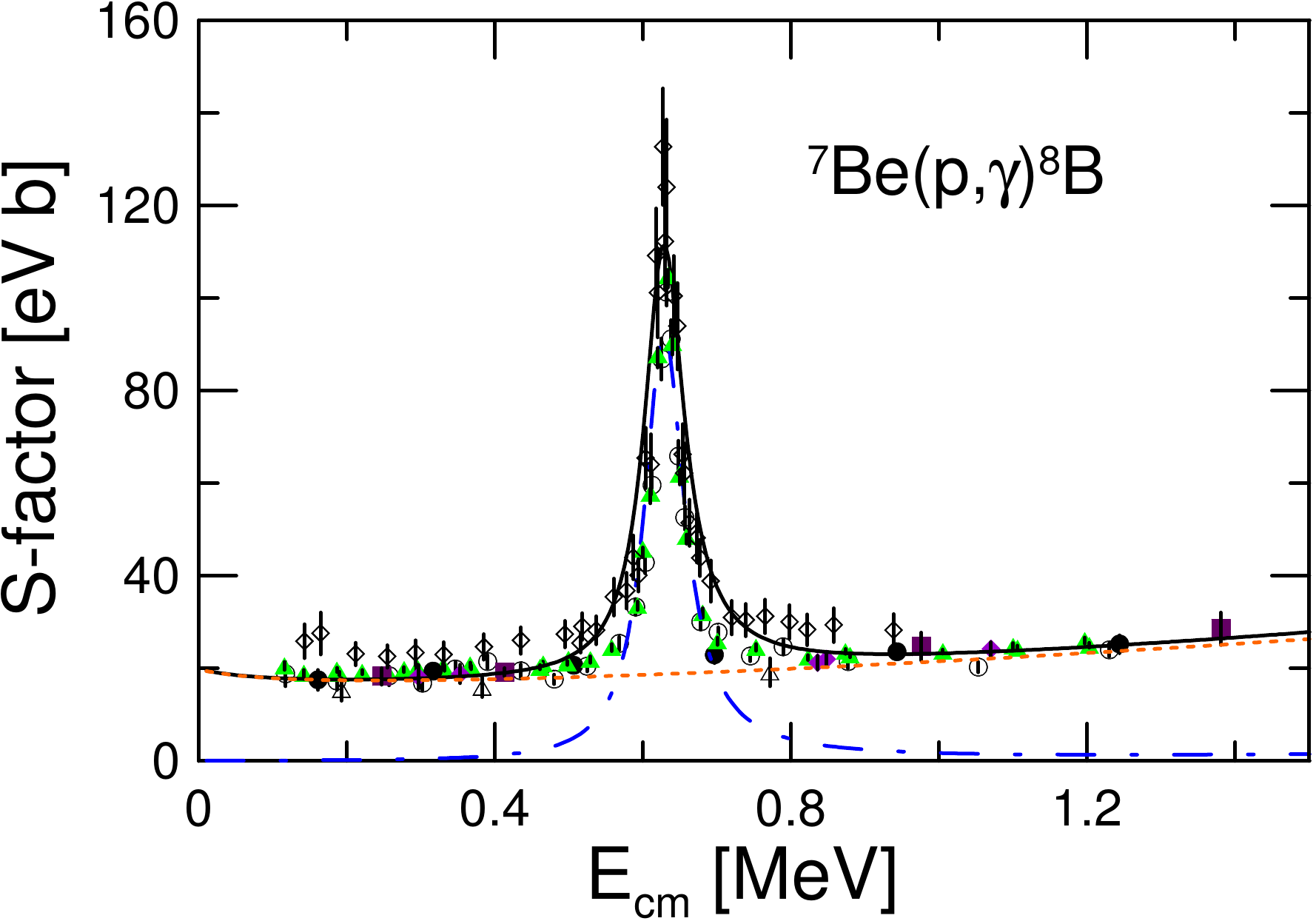}
\caption{{\it Top} - Potential model calculation for the
reaction $^{7}${\rm Li} ({\rm p}, $\gamma$)$^{8}${\rm Be}.
Experimental data are from  Ref.
\cite{Zahnow95}. 
{\it Bottom} - Single-particle model calculations for the reaction $^{7}${\rm Be}%
$(p,\gamma)^{8}{\rm B}$. The dashed-dotted line is the calculation
for the M1 resonance at $E_{cm}=0.63$ MeV and the dotted line is for
the non-resonant capture. Experimental data are from Refs.
\cite{VCK70, FED83, Baby03, Junghans03, Iwasa99,Kav69}. The total S
factor is shown as a
solid line. }
\end{figure}

In the calculation of $\sigma_{L,J_{b}}^{{\rm d.c.}}$ above, one often
meets the situation in which only the asymptotic part of $I_{b}(r)$ and
$\Psi_{c}(r)$ contributes significantly to the integral over $r$. In these
situations, $u_{c}(r)$ is also well described by a simple two-body
scattering wave (e.g. Coulomb waves). Therefore the radial integration in
$\sigma_{L,J_{b}}^{{\rm d.c.}}$ can be done accurately and the only
remaining information from the many-body physics at short-distances is
contained in the asymptotic normalization coefficient $C_{i}$, i.e.
$\sigma_{L,J_{b}}^{{\rm d.c.}}\propto C_{i}^{2}$. We thus run into an
effective theory for radiative capture cross sections, in which the constants
$C_{i}$ carry all the information about the short-distance physics, where the
many-body aspects are relevant. It is worthwhile to mention that these
arguments are reasonable for proton capture at very low energies, because of
the Coulomb barrier.

As the overlap integral,   Eq. \ref{whitt}, asymptotically
becomes a Whittaker function, so does the single particle
bound-state wavefunction $u_{\alpha}$. If we call the single particle ANC by $b_{i}$, then the
relation between the ANC obtained from experiment, or a microscopic
model, with the single particle ANC is given by $(SF)_{i}b_{i}^{2}%
=C_{i}^{2}$. This becomes clear from     Eq. \ref{SFS}. The values of
$(SF)_{i}$ and $b_{i}$ obtained with the simple potential model are
useful telltales of the complex short-range many-body physics of
radiative capture reactions \cite{HBG10}. 

Many reactions of interest for nuclear astrophysics involve nuclei
close to the dripline. To describe these reactions, a knowledge of
the structure in the continuum is a crucial feature.  One
basic theoretical problem is to what extent we know the form of the
effective interactions for threshold states. It is also hopeless
that these methods can be accurate in describing high-lying states
in the continuum. In particular, it is not worthwhile to pursue this
approach to describe direct nuclear reactions.

\subsection{Resonating group method}

One immediate goal can be achieved in the coming years by using the
Resonating Group Method (RGM) or the
Generator Coordinate Method (GCM). These are a set of coupled integro-differential
equations of the form
\begin{equation}
\sum_{\alpha'} \int d^3 r'
\left[
H^{AB}_{\alpha\alpha'}({\bf r,r'})-EN^{AB}_{\alpha\alpha'}({\bf r,r'})
\right]
g_{\alpha'}({\bf r'})=0,\label{RGM}
\end{equation}
where $H^{AB}_{\alpha\alpha'}({\bf r,r'})=\langle \Psi_A(\alpha,{\bf r})|H|
\Psi_B(\alpha',{\bf r'}) \rangle$ and $N^{AB}_{\alpha\alpha'}({\bf r,r'})
=\langle \Psi_A(\alpha,{\bf r})|
\Psi_B(\alpha',{\bf r'}) \rangle$. In these equations $H$ is the Hamiltonian for the
system of two nuclei (A and B) with the energy $E$, $\Psi_{A,B}$ is the wavefunction
of nucleus A (and B), and $g_{\alpha}({\bf r})$ is a function to be found by numerical
solution of Eq. \ref{RGM}, which describes the relative motion of A and B in channel
$\alpha$.
Full antisymmetrization between nucleons of A and B are implicit.
Modern nuclear shell-model calculations, including the No-Core-Shell-Model (NCSM) are able to
provide the wavefunctions $\Psi_{A,B}$ for light nuclei \cite{QN08}. But the Hamiltonian involves
an effective interaction in the continuum between the clusters A and B. 
Overlap integrals of the type $I_{Aa}(r)=\langle
\Psi_{A-a}|\Psi_A\rangle$ for bound states has been calculated  within the NCSM. This is one of
the inputs necessary to calculate S-factors for radiative capture,
$S_\alpha \sim |\langle g_{\alpha}|{\cal O}_{EM}|I_{Aa}\rangle|^2$, where
${\cal O}_{EM}$ is a corresponding electromagnetic operator. The left-hand
side of this equation is to be obtained by solving Eq. \ref{RGM}.
For some cases, in particular for the p$+^7$Be reaction, the
distortion caused by the microscopic structure of the cluster does
not seem to be crucial to obtain the wavefunction in the continuum.
The wavefunction is often obtained by means of a potential model.
The NCSM overlap integrals, $I_{Aa}$, can also be corrected to
reproduce the right asymptotics \cite{NBC05,PRQ11}, given by
$I_{Aa}(r)\propto W_{-\eta,l+1/2}(2k_0r)$, where $\eta$ is the
Sommerfeld parameter, $l$ the angular momentum, $k_0=\sqrt{2\mu
E_0}/\hbar$ with $\mu$ the reduced mass and $E_0$ the separation
energy.

\begin{figure}
\begin{center}\includegraphics[
height=2.5 in]{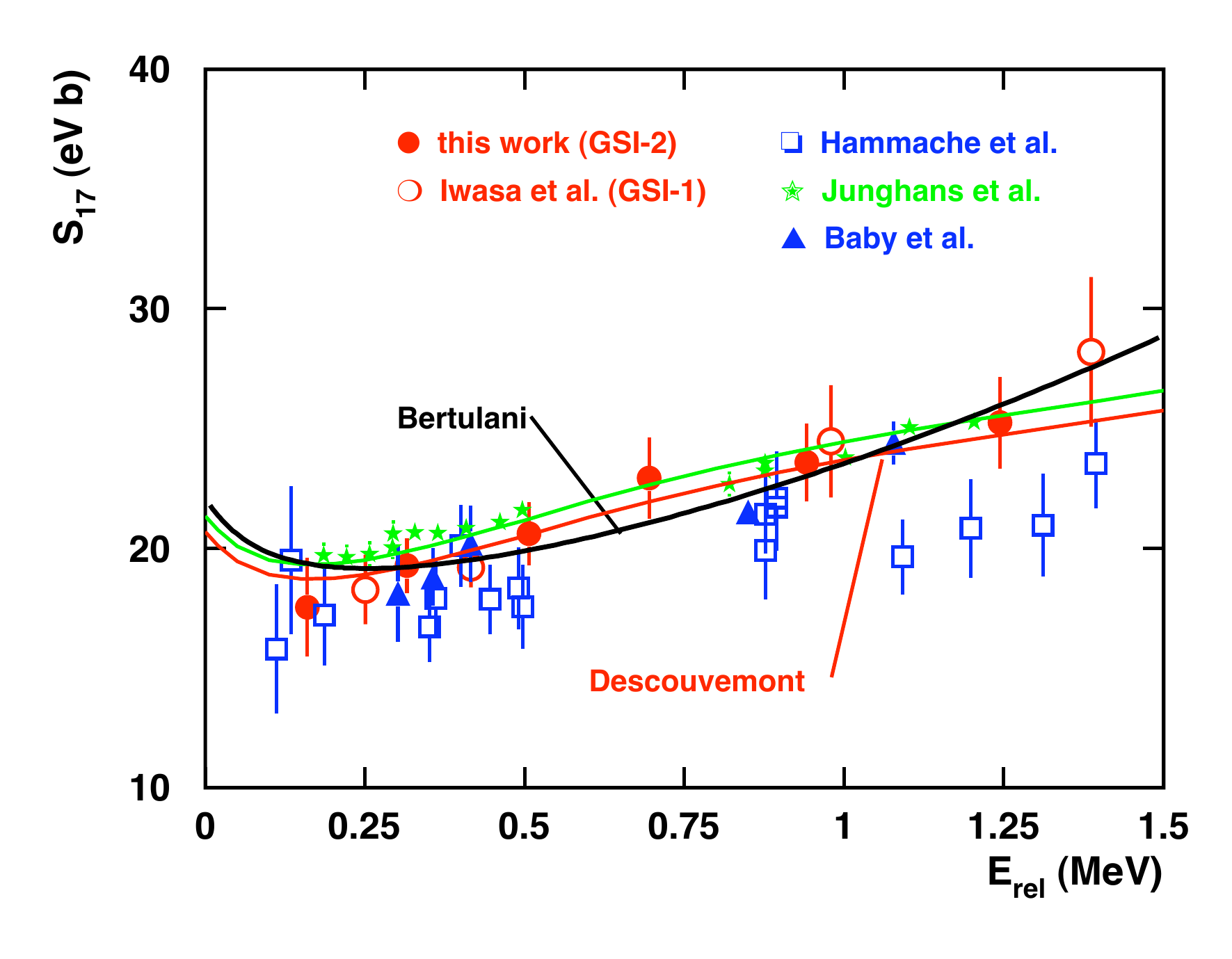}
\caption{ World data on $^7$Be(p,$\gamma$)$^8$B compared to theoretical calculations. }
\label{s17ber}
\end{center}
\end{figure}

A step in the direction of reconciling structure and reactions for
the practical purpose of obtaining astrophysical S-factors, along
the lines described in the previous paragraph, was obtained in Ref.
\cite{NBC05}. The wavefunctions obtained in this way were
shown to reproduce very well the momentum distributions in knockout
reactions of the type $^8$B$+A\longrightarrow \ ^7$Be$+X$. The astrophysical S-factor
for the reaction  $^{7}$Be$($p$,\gamma)^{8}$B was also calculated
and excellent agreement was found with the experimental data in both
direct and indirect measurements \cite{NBC05,PRQ11}. The low- and
high-energy slopes of the  S-factor obtained with the NCSM is well
described by the fit
\begin{equation}
S_{17}(E)=(22.109\ {\rm eV.b}){1+5.30E+1.65E^2+0.857E^3 \over 1+E/0.1375}  ,
\end{equation}
where E is the relative energy (in MeV) of p$+^7$Be in their
center-of-mass. This equation corresponds to a Pad\'e approximant of
the S-factor. A subthreshold pole due to the binding energy of $^8$B
is responsible for the denominator \cite{JKS98,WK81}. Figure \ref{s17ber} show the world data on $^7$Be(p,$\gamma$)$^8$B compared to a few of the theoretical calculations. The recent compilation published in Ref. \cite{RMP11} recommends $S_{17} = 20.8 \pm 0.7 \ {\rm (expt)} \pm 1.4\ {\rm (theor)}$ eV b.

\section{Direct reactions and the role of radioactive beams}

\subsection{Transfer reactions}

Transfer reactions $A(a,b)B$ are effective when a momentum matching
exists between the transferred particle and the internal particles
in the nucleus. Thus, beam energies should be in the range of a few
10 MeV per nucleon. Low energy reactions of
astrophysical interest can be extracted directly from breakup
reactions $A+a \longrightarrow b+c+B$ by means of the  Trojan
Horse method (THM)  \cite{Bau86}. If the
Fermi momentum of the particle $x$ inside $a=(b+x)$ compensates for
the initial projectile velocity $v_a$, the low energy reaction
$A+x=B+c$ is induced at very low (even vanishing) relative energy
between $A$ and $x$. To
show this, one writes the DWBA cross section for the breakup reaction as 
$$
{d^3\sigma\over d\Omega_b d\Omega_c dE_b} \propto \left |\sum_{lm} T_{lm}({\bf k_a,
k_b, k_c}) S_{lx} Y_{lm}({\bf k_c})\right|^2,$$ where $$T_{lm}=
<\chi_b^{(-)} Y_{lm} f_l |V_{bx}|\chi_a^{+}\phi_{bx}>.$$ The
threshold behavior $E_x$ for the breakup cross section
$\sigma_{A+x\rightarrow B+c} =(\pi/k_x^2)\sum_l (2l+1)|S_{lx}|^2$ is
well known: since $|S_{lx}|\sim \exp(-2\pi \eta)$, then
$\sigma_{A+x\rightarrow B+c}\sim (1/k_x^2)\ \exp(-2\pi \eta)$. In
addition to the threshold behavior of $S_{lx}$, the breakup cross
section is also governed by the threshold behavior of $f_l(r)$,
which for $r\longrightarrow \infty$ is given by $f_{l_x}\sim
(k_xr)^{1/2} \ \exp(\pi \eta ) \ K_{2l+1}(\xi)$, where $K_l$ denotes
the Bessel function of the second kind of imaginary argument. The
quantity $\xi$ is independent of $k_x$ and is given by
$\xi=(8r/a_B)^{1/2}$, where $a_B=\hbar^2/mZ_AZ_xe^2$ is the Bohr
length. From this one obtains that $(d^3/d\Omega_b d\Omega_c dE_b)
({E_x \rightarrow 0}) \approx {\rm const.}$. The coincidence cross
section tends to a constant which will in general be different from
zero. This is in striking contrast to the threshold behavior of the
two particle reaction $A+x=B+c$. The strong barrier penetration
effect on the charged particle reaction cross section is canceled
completely by the behavior of the factor $T_{lm}$ for $\eta
\rightarrow \infty$. Basically, this technique extends the method of
transfer reactions to continuum states. very successful results
using this technique have been reported \cite{Con07,Piz11}.

Another transfer method, coined as Asymptotic Normalization
Coefficient (ANC) technique \cite{Muk90,Tri06,Mu10} relies on fact that the amplitude for
the radiative capture cross section $b+x\longrightarrow a+ \gamma$
is given by $$M=<I_{bx}^a({\bf r_{bx}})|{\cal O}({\bf r_{bx}})|
\psi_i^{(+)}({\bf r_{bx}})>,$$ where $$I_{bx}^a=<\phi_a(\xi_b, \ \xi_x,\ {\bf %
r_{bx}}) |\phi_x(\xi_x)\phi_b(\xi_b)>$$ is the integration over the
internal coordinates $\xi_b$, and $\xi_x$, of $b$ and $x$,
respectively. For low energies, the overlap integral $I_{bx}^a$ is
dominated by contributions from large $r_{bx}$. Thus, what matters
for the calculation of the matrix element $M$ is the asymptotic
value of $I_{bx}^a\sim C_{bx}^a \ W_{-\eta_a, 1/2}(2\kappa_{bx}
r_{bx})/r_{bx}$, where $C_{bx}^a$ is the ANC and $W$ is the
Whittaker function. This coefficient is the product of the
spectroscopic factor and a normalization constant which depends on
the details of the wave function in the interior part of the
potential. Thus, $C_{bx}^a$ is the only unknown factor needed to
calculate the direct capture cross section. These normalization
coefficients can be found from: 1) analysis of classical nuclear
reactions such as elastic scattering [by extrapolation of the
experimental scattering phase shifts to the bound state pole in the
energy plane], or 2) peripheral transfer reactions whose amplitudes
contain the same overlap function as the amplitude of the
corresponding astrophysical radiative capture cross section. 

To illustrate this technique, let us consider the proton transfer reaction $%
A(a,b)B$, where $a=b+p$, $B=A+p$. Using the asymptotic form of the
overlap integral the DWBA cross section is given by $$d\sigma/d\Omega
=
\sum_{J_Bj_a}\left [{(C_{Ap}^a)^2\over \beta^2_{Ap}}\right]\left [{(C_{bp}^a)^2\over \beta^2_{bp}}\right]
{\tilde \sigma}$$ where $\tilde \sigma$ is the reduced cross section
not depending on the nuclear structure, $\beta_{bp}$ ($\beta_{Ap}$)
are the asymptotic normalization of the shell model bound state
proton wave functions in nucleus $a (B)$ which are related to the
corresponding ANC's of the overlap function as $(C_{bp}^a)^2
=S^a_{bp} \beta^2_{bp}$. Here $S^a_{bp}$ is the spectroscopic
factor. Suppose the reaction $A(a,b)B$ is peripheral. Then each of
the bound state wave functions entering $\tilde \sigma$ can be
approximated by its asymptotic form and $\tilde \sigma \propto
\beta_{Ap}^2 \beta_{bp}^2$. Hence $d\sigma/d\Omega =
\sum_{j_i}(C_{Ap}^a)^2(C_{bp}^a)^2 R_{Ba}$ where $R_{Ba}={\tilde
\sigma}/\beta^2_{Ap} \beta^2_{bp}$ is independent of $\beta^2_{Ap}$
and $\beta^2_{bp}$. Thus for surface reactions the DWBA cross
section is actually parameterized in terms of the product of
the square of the ANC's of the initial and the final nuclei $%
(C_{Ap}^a)^2(C_{bp}^a)^2$ rather than spectroscopic factors. This
effectively removes the sensitivity in the extracted parameters to
the internal structure of the nucleus.
One of the many advantages of using transfer reaction techniques
over direct measurements is to avoid the treatment of the screening
problem \cite{Con07,Mu10}.

But do we really understand transfer reactions well enough? Let us take an example from literature \cite{Corradi}. 
Assuming that $\alpha$ in Eq.  \ref{cc} is simply the time $t$, and using the first-Born approximation (i.e, taking $a_k\sim a_0\delta_{k0}$), the amplitude to excite the channel $\phi_k$ from an initial channel $\phi_0$ is given by $a_k=-i\hbar \int \langle \phi_0 |U|\phi_k\rangle \exp[i(E_k-E_0)t/\hbar]$. The Born approximation can be applied to transfer reactions. The probability to transfer a nucleon in nucleus A from channel $\alpha$ to a nucleon in nucleus B  in channel $\beta$ is given by
 \begin{equation}
P_{\beta\alpha}\sim \left| -i\hbar \int_{-\infty}^\infty dt F_{\beta\alpha}({\bf R}) \exp\left[i{E_\beta-E_\alpha)t\over \hbar}+(\cdots)\right]\right|^2, \label{transf}
 \end{equation} 
 where ${\bf R}$ is the nucleus-nucleus distance and $F_{\beta\alpha}({\bf R})$ is the from factor given by
  \begin{equation}
F_{\beta\alpha}({\bf R})=  \int d^3r e^{i{\bf Q}\cdot {\bf r}} \phi_\beta({\bf R}+{\bf r})\left( V_1-\langle U\rangle \right) \phi_\alpha({\bf r}),  \label{transf2}
\end{equation} 
where ${\bf Q}$ is the momentum transfer in the reaction, $U$ is the total (optical) potential, and $V_1$ is the potential of the nucleon with one of the nuclei. Why not $V_2$? In the literature, using $V_1$ ($V_2$) goes by the name  ``prior" (``post)-form. It has been shown in the past that the post and prior forms of breakup and transfer reactions lead to the same result.  

In figure \ref{fig:1} (bottom) one sees the probabilities for multi-nucleon transfer in $^{96}$Zr+$^{40}$Ca, as a function of the closest approach distance $D=(Z_1Z_2e^2/2E)  [1+1/\sin (\theta/2)]$. Transfer is most likely to occur when the nuclei are at their closest point, $D$. The tunneling probability depends exponentially on this distance, $P_{tr}/\sin(\theta/2)\sim \exp(-2\alpha D)$. This approximation arises from Eqs. \ref{transf} and \ref{transf2}. If one neglects correlations, two-nucleon transfer probabilities are given in terms one-nucleon transfer probabilities: $P_{2n} = \left( P_{1n}\right)^2$. For three-nucleon transfer $P_{3n} = P_{1n}P_{2n}$, and so on. These are shown by the straight lines in figure \ref{fig:1} (Right). All seems to work well, except that one needs an enhancement of a factor 3 to get $P_{2n}$ from theory \cite{Ku90}. That is what happens when theorists do not know what to do \cite{Corradi}.

\subsection{Intermediate energy Coulomb excitation}

At low-energies,
 the theory of Coulomb excitation is very well understood \cite{AW75}. A large
number of small corrections are now well known in the theory and are
necessary in order to analyze experiments on multiple excitation and
reorientation effects. At the other end, the Coulomb excitation of
relativistic heavy ions is characterized by straight-line
trajectories with impact parameter $b$ larger than the sum of the
radii of the two colliding nuclei \cite{WA79}. It was also shown that
a quantum theory for relativistic Coulomb excitation leads to
modifications of the semiclassical results~\cite{Ber88}. In
Refs.~\cite{AB89,BN93} the inclusion of relativistic effects in
semiclassical and quantum formulations of Coulomb excitation was
fully clarified.

Recently, the importance of relativistic effects in Coulomb
excitation of a projectile by a target with charge $Z_{2}$, followed
by gamma-decay, in nuclear reactions at intermediate energies was
studied in details. The Coulomb excitation cross section is given by
\begin{eqnarray}
{\frac{d\sigma_{i\rightarrow f}}{d\Omega}}&=&\left(
\frac{d\sigma}{d\Omega }\right)
_{\mathrm{el}}\frac{16\pi^{2}Z_{2}^{2}e^{2}}{\hbar^{2}}\nonumber \\ &\times& \sum
_{\pi\lambda\mu}{\frac{B(\pi\lambda,I_{i}\rightarrow
I_{f})}{(2\lambda
+1)^{3}}}\mid S(\pi\lambda,\mu)\mid^{2},\label{cross_2}%
\end{eqnarray}
where $B(\pi\lambda,I_{i}\rightarrow I_{f})$ is the reduced
transition probability of the projectile nucleus, $\pi\lambda=E1,\
E2,$ $M1,\ldots$ is the multipolarity of the excitation, and
$\mu=-\lambda,-\lambda+1,\ldots,\lambda$.

The relativistic corrections
to the Rutherford formula for $\left(
d\sigma /d\Omega\right)  _{\mathrm{el}}$ has been investigated
in Ref. \cite{AAB90}.
It was shown that
the scattering angle increases by up to 6\% when relativistic
corrections are included in nuclear
collisions at 100 MeV/nucleon. The effect on the elastic scattering
cross section is even more drastic: up to $13\%$ for center-of-mass
scattering angles around 0-4 degrees.

The orbital integrals
$S(\pi\lambda,\mu$) contain the information about relativistic
corrections. Inclusion of absorption effects in
$S(\pi\lambda,\mu$) due to the imaginary part of an optical
nucleus-nucleus potential where worked out in Ref. \cite{BN93}.
These orbital integrals depend on the Lorentz factor
$\gamma=(1-v^{2}/c^{2})^{-1/2}$, with $c$ being the speed of light,
on the multipolarity $\pi\lambda\mu$, and on the adiabacity
parameter $\xi (b)=\omega_{fi}b/\gamma v<1$, where
$\omega_{fi}=\left( E_{f}-E_{i}\right) /\hbar$ is the excitation
energy (in units of $\hbar$) and $b$ is the impact parameter.

Ref. \cite{Ber03} has shown that at 10 MeV/nucleon the
relativistic corrections are important only at the level of 1\%. At
500 MeV/nucleon, the correct treatment of the recoil corrections  is
relevant on the level of 1\%. Thus the non-relativistic treatment of
Coulomb excitation~\cite{AW75} can be safely used for energies below
about 10 MeV/nucleon and the relativistic treatment with a
straight-line trajectory~\cite{WA79} is adequate above about 500
MeV/nucleon. However at energies around 50 to 100 MeV/nucleon,
accelerator energies common to most radioactive beam facilities, it is very important to use a correct
treatment of recoil and relativistic effects, both kinematically and
dynamically.
At these energies, the corrections can add up to 50\%.
These effects were also shown in Ref.~\cite{AB89} for the case of
excitation of giant resonances in collisions at intermediate
energies.

A reliable extraction of useful nuclear properties,
like the electromagnetic response (B(E2)-values, $\gamma$-ray
angular distribution, etc.) from Coulomb excitation experiments at
intermediate energies requires a proper treatment of special
relativity \cite{Ber03,BCG03}. The dynamical relativistic effects have often been
neglected in the analysis of experiments elsewhere (see, e.g. \cite{Glas01}).
The effect is highly non-linear, i.e. a 10\% increase in
the velocity might lead to a 50\% increase (or decrease) of certain
physical observables. A general review of the importance of the relativistic
dynamical effects in intermediate energy collisions has been presented in
Ref. \cite{Ber05_work,Ber05}.

\subsection{The Coulomb dissociation method}

The Coulomb dissociation method is quite simple.
The (differential, or angle integrated) Coulomb breakup cross section
for $a+A\longrightarrow b+c+A$ follows from Eq. \ref{cross_2}.
It can be rewritten as
\begin{equation}
{d\sigma_{C}^{\pi\lambda
}(\omega)\over d\Omega}=F^{\pi\lambda}(\omega;\theta;\phi)\ .\
\sigma_{\gamma+a\ \rightarrow\ b+c}^{\pi\lambda}(\omega),\label{CDmeth}
\end{equation}
where $\omega$ is the energy transferred from the relative motion to the
breakup, and $\sigma_{\gamma+a\ \rightarrow\ b+c}^{\pi\lambda}(\omega)$ is the photo nuclear cross
section for the multipolarity ${\pi\lambda}$ and photon energy $\omega$. The
function $F^{\pi\lambda}$ depends on $\omega$, the relative motion energy,
nuclear charges and radii, and the scattering angle $\Omega=(\theta,\phi)$.
$F^{\pi\lambda}$
can be reliably calculated \cite{Ber88} for each
multipolarity ${\pi\lambda}$. Time reversal allows one to deduce the radiative
capture cross section $b+c\longrightarrow a+\gamma$ from $\sigma_{\gamma+a\ \rightarrow\ b+c}%
^{\pi\lambda}(\omega)$. This method was proposed in Ref. \cite{BBR86} and has
been tested successfully in a number of reactions of interest for astrophysics.
The most celebrated case is the
reaction $^{7}$Be$($p$,\gamma)^{8}$B \cite{Tohru}, followed by numerous
experiments in the last decade (see e.g. Ref. \cite{EBS05}).

Eq. \ref{CDmeth} is based on first-order perturbation theory. It
also assumes that the nuclear contribution to the breakup is small,
or that it can be separated under certain experimental conditions.
The contribution of the nuclear breakup has been examined by several
authors (see, e.g. \cite{BG98}). $^8$B has a small proton separation
energy ($\approx 140$ keV). For such loosely-bound systems it had
been shown that multiple-step, or higher-order effects, are
important \cite{BC92}. These effects occur by means of
continuum-continuum transitions. Detailed studies of dynamic
contributions to the breakup were explored in refs.
\cite{BBK92,BB93} and in several other publications which followed.
The role of higher multipolarities (e.g., E2 contributions
\cite{Ber94,GB95,EB96} in the reaction $^{7}$Be$($p$,\gamma)^{8}$B)
and the coupling to high-lying states has also to be investigated
carefully. It has also been shown that the
influence of giant resonance states is small \cite{Ber02}. 

\subsection{Charge exchange reactions}

During core collapse, temperatures and densities are high enough to
ensure that nuclear statistical equilibrium  is achieved. This
means that for sufficiently low entropies, the matter composition is
dominated by the nuclei with the highest binding energy for a given
$Y_{e}$. Electron capture reduces $Y_{e}$, driving the nuclear
composition to more neutron rich and heavier nuclei, including those
with $N>40$, which dominate the matter composition for densities
larger than a few $10^{10}$~g~cm$^{-3}$. As a consequence of the
model applied in collapse simulations, electron capture on nuclei
ceases at these densities and the capture is entirely due to free
protons. To understand the whole process it is necessary to obtain
Gamow-Teller matrix elements which are not accessible in beta-decay
experiments. Many-body theoretical calculations are right now the
only way to obtain the required matrix elements. This situation can
be remedied experimentally by using charge-exchange reactions.
Charge exchange reactions induced in (p,n) reactions are often used to obtain
values of Gamow-Teller matrix elements, $B(GT)$, which cannot be extracted from
beta-decay experiments. This approach relies on the similarity in spin-isospin
space of charge-exchange reactions and $\beta$-decay operators. As a result of
this similarity, the cross section $\sigma($p,\ n$)$ at small momentum
transfer $q$ is closely proportional to $B(GT)$ for strong transitions
\cite{Tad87},
\begin{equation}
{d\sigma\over dq}(q=0)=KN_D|J_{\sigma\tau}|^2 B(\alpha),\label{Tad87}
\end{equation}
where $K$ is a kinematical factor, $N_D$ is a distortion factor (accounting for
initial and final state interactions), $J_{\sigma\tau}$ is the Fourier transform
of the effective nucleon-nucleon interaction, and $B(\alpha=F,GT)$ is the reduced transition
probability for non-spin-flip, $$B(F)=
(2J_i+1)^{-1}| \langle f ||\sum_k  \tau_k^{(\pm)} || i \rangle |^2,$$
and spin-flip,
$$B(GT)=
(2J_i+1)^{-1}| \langle f ||\sum_k \sigma_k \tau_k^{(\pm)} || i \rangle |^2,$$ transitions.

Eq. \ref{Tad87}, valid for one-step processes, was proven to work rather well
for (p,n) reactions (with
a few exceptions). For heavy ion reactions the formula might not work so well.
This has been investigated in refs. \cite{Len89,Ber93,BL97}. In Ref. \cite{Len89} it
was shown that multistep processes involving the physical exchange of a proton and
a neutron can still play an important role up to bombarding energies of 100 MeV/nucleon.
Refs. \cite{Ber93,BL97} use the isospin terms of the effective interaction to show that
deviations from the Taddeucci formula are common under many circumstances.
As shown in Ref. \cite{Aus94}, for important GT transitions
whose strength are a small fraction of the sum rule the direct relationship
between $\sigma($p,\ n$)$ and $B(GT)$ values also fails to exist. Similar
discrepancies have been observed \cite{Wat85} for reactions on some odd-A
nuclei including $^{13}$C, $^{15}$N, $^{35}$Cl, and $^{39}$K and for
charge-exchange induced by heavy ions \cite{BL97,St96}.
Undoubtedly, charge-exchange reactions such as (p,n), ($^{3}$He,t)
and heavy-ion reactions (A,A$\pm$1) can provide information on the
$B(F)$ and $B(GT)$ values needed for astrophysical purposes \cite{Fu09}.

\subsection{Knock-out reactions}

Exotic nuclei are the raw
materials for the synthesis of the heavier
elements in the Universe, and are of considerable
importance in nuclear astrophysics. Modern
shell-model calculations are also now able to
include the effects of residual interactions
between pairs of
nucleons, using forces that reproduce the
measured masses, charge radii and low-lying
excited states of a large number of nuclei.
For very exotic nuclei the small additional
stability that comes with the filling of a particular
orbital can have profound effects upon their
existence as bound systems, their lifetimes and
structures. Thus, verifications of the ordering,
spacing and the occupancy of orbitals are
essential in assessing how exotic nuclei evolve in
the presence of large neutron or proton
imbalance and our ability to predict these
theoretically. Such spectroscopy of the states of
individual nucleons in short-lived nuclei uses
direct nuclear reactions.  

The early interest in knockout reactions came from studies of
nuclear halo states, for which the narrow momentum distributions
of the core fragments
in a qualitative way revealed the large spatial extension of the
halo wave function. It was shown 
\cite{ber92} that the longitudinal component of the momentum
(taken along the beam or $z$ direction) gave the most accurate
information on the intrinsic properties of the halo and that it
was insensitive to details of the collision and the size of the
target. In contrast to this, the transverse distributions of the
core are significantly broadened by diffractive effects and by
Coulomb scattering. For experiments that observe the nucleon
produced in elastic breakup, the transverse momentum is entirely
dominated by diffractive effects, as illustrated \cite{ann94} by
the angular distribution of the neutrons from the reaction
$^{9}$Be($^{11}$Be,$^{10}$Be+n)X. In this case, the width of the
transverse momentum distribution reflects essentially the size of
the target  \cite{BH04}.

Most practical studies of medium corrections in nucleon-nucleon scattering are carried out by considering  the effective two-nucleon  interaction in infinite  nuclear matter. This is known as the G-matrix method, an is obtained from a solution of the Bethe-Goldstone equation 
\begin{eqnarray}
&&\langle\mathbf{k}|\mathrm{G}(\mathbf{P},\rho_1,\rho_2)|\mathbf{k}_{0}\rangle
=\langle\mathbf{k}|\mathrm{v}_{NN}|\mathbf{k}_{0}\rangle-
\nonumber \\
&&\int{\frac
{d^{3}k^{\prime}}{(2\pi)^{3}}}{\frac{\langle\mathbf{k}|\mathrm{v}%
_{NN}|\mathbf{k^{\prime}}\rangle Q(\mathbf{k^{\prime}},\mathbf{P}%
,\rho_1,\rho_2)\langle\mathbf{k^{\prime}}|\mathrm{G}(\mathbf{P},\rho_1,\rho_2)|\mathbf{k}%
_{0}\rangle}{E(\mathbf{P},\mathbf{k^{\prime}})-E_{0}-i\epsilon}},\nonumber \\
\label{10}%
\end{eqnarray}
with $\mathbf{k}_{0}$, $\mathbf{k}$, and $\mathbf{k^{\prime}}$ the
initial, final, and intermediate relative momenta of the NN pair,
${\bf k}=({\bf k}_1-{\bf k}_2)/2$ and ${\bf P}=({\bf k}_1+{\bf k}_2)/2$. If energy and momentum is conserved in the binary collision,
${\bf P}$ is conserved in magnitude and direction, and the magnitude of ${\bf k}$ is also conserved. $\mathrm{v}_{NN}$ is the
nucleon-nucleon potential. $E$ is the energy of the two-nucleon
system, and $E_{0}$ is the same quantity on-shell. Thus
$
E(\mathbf{P},\mathbf{k})=e(\mathbf{P}+\mathbf{k})+e(\mathbf{P}-\mathbf{k}%
)$, with $e$ the single-particle energy in nuclear matter. It is also implicit in  Eq. \ref{10} that the final momenta ${\bf k}$ of the NN-pair
 also lie outside the  range of occupied states. 

\begin{figure}[h]
\begin{center}
\includegraphics[height=2.5 in]{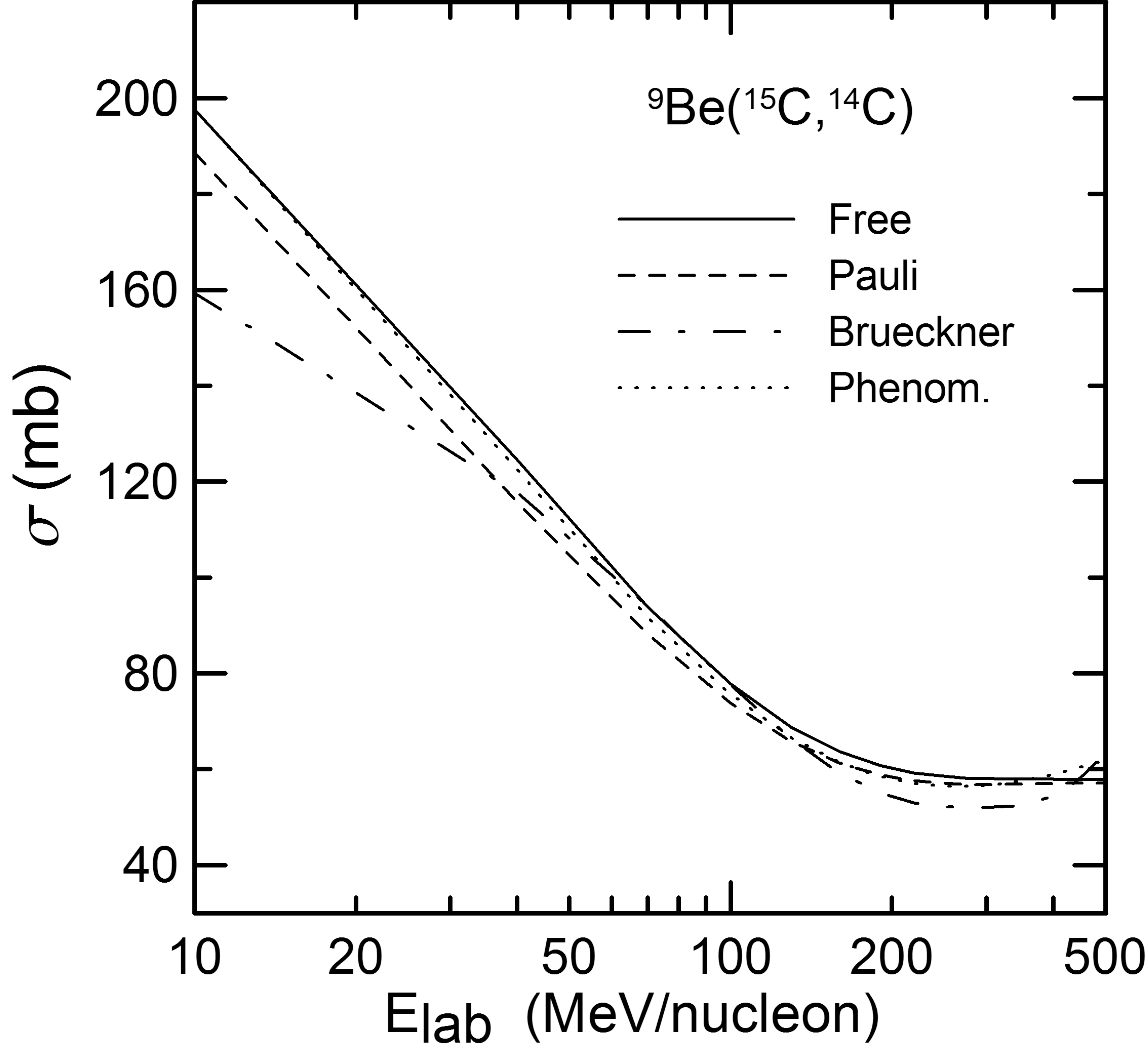}
\caption{Total  knockout cross sections for removing the $l=0$ halo neutron of $^{15}$C, bound by 1.218 MeV, in the reaction $^9$Be($^{15}$C,$^{14}$C$_{gs}$). The solid curve is obtained with the use of free nucleon-nucleon cross sections.  The dashed curve includes the geometrical effects of Pauli blocking. The dashed-dotted curve is the result using the Brueckner theory,  and the dotted curve is a  phenomenological parametrization.}
\end{center}
\label{sko15c}
\end{figure}

In Ref. \cite{BC10}  the numerical calculations have been performed to account for the geometric effect of Pauli blocking. A parametrization has been devised which fits the numerical results. The parametrization reads 
\begin{eqnarray}
\sigma_{NN}(E,\rho_1,\rho_2) &=&\sigma_{NN}^{free}(E)\nonumber \\
&\times& {1 \over 1+1.892{\left({2\rho_<\over \rho_0}\right)}\left({{|\rho_1-\rho_2|\over \tilde{\rho}\rho_0}}\right)^{2.75}}\nonumber \\
&\times& 
\left\{
\begin{array}
[c]{c}%
{1-{37.02 \tilde{\rho}^{2/3}\over E}}, \ \ \   {\rm if} \ \ E>46.27 \tilde{\rho}^{2/3}\\ \, \\
{{E\over 231.38\tilde{\rho}^{2/3}}},\ \ \ \ \  {\rm if} \ \ E\le 46.27 \tilde{\rho}^{2/3}\end{array}
\right.\nonumber \\ &&
\label{VM1}
\end{eqnarray}
where $E$ is the laboratory energy in MeV, $\rho_i$ is the local density of nucleus $i$, $\rho_<={\rm min} (\rho_1,\rho_2)$ and $\tilde{\rho}=(\rho_1+\rho_2)/\rho_0$,  with $\rho_0=0.17$ fm$^{-3}$.

The Brueckner method goes beyond a treatment of Pauli blocking, and has been presented in several works, e.g.  in Ref. \cite{LM:1993,LM:1994}, where a simple parametrization was given, which we will from now on refer as Brueckner theory. It reads (the misprinted factor 0.0256 in Ref. \cite{LM:1994} has been corrected to 0.00256)
\begin{eqnarray}
\sigma_{np}  &  =& \left[ 31.5 +0.092\left| 20.2-E^{0.53}\right|^{2.9}\right] {1+0.0034E^{1.51} \rho^2\over 1+21.55\rho^{1.34}} \nonumber\\
\sigma_{pp}  &  = &\left[ 23.5 +0.00256\left( 18.2-E^{0.5}\right)^{4.0}\right] {1+0.1667E^{1.05} \rho^3\over 1+9.704\rho^{1.2}}  \nonumber \\ \label{brueckner}
\end{eqnarray}

A modification of the above parametrization was done in Ref. \cite{Xian98}, which consisted in combining the free nucleon nucleon cross sections parametrized in Ref. \cite{Cha90} with the Brueckner theory results of Ref. \cite{LM:1993,LM:1994}. 

To test the influence of the medium effects in nucleon knockout reactions, we consider the removal of the $l=0$ halo neutron of $^{15}$C, bound by 1.218 MeV,  and the $l=0$ neutron knockout from $^{34}$Ar, bound by 17.06 MeV. The reaction studied is $^9$Be($^{15}$C,$^{14}$C$_{gs}$). The total cross sections as a function of the bombarding energy are shown in figures \ref{sko15c}. The solid curve is obtained with the use of free nucleon-nucleon cross sections.  The dashed curve includes the geometrical effects of Pauli blocking. The dashed-dotted curve is the result using the Brueckner theory,  and the dotted curve is the phenomenological parametrization of the free cross section.

\begin{figure}[h]
\begin{center}
\includegraphics[height=2.5 in]{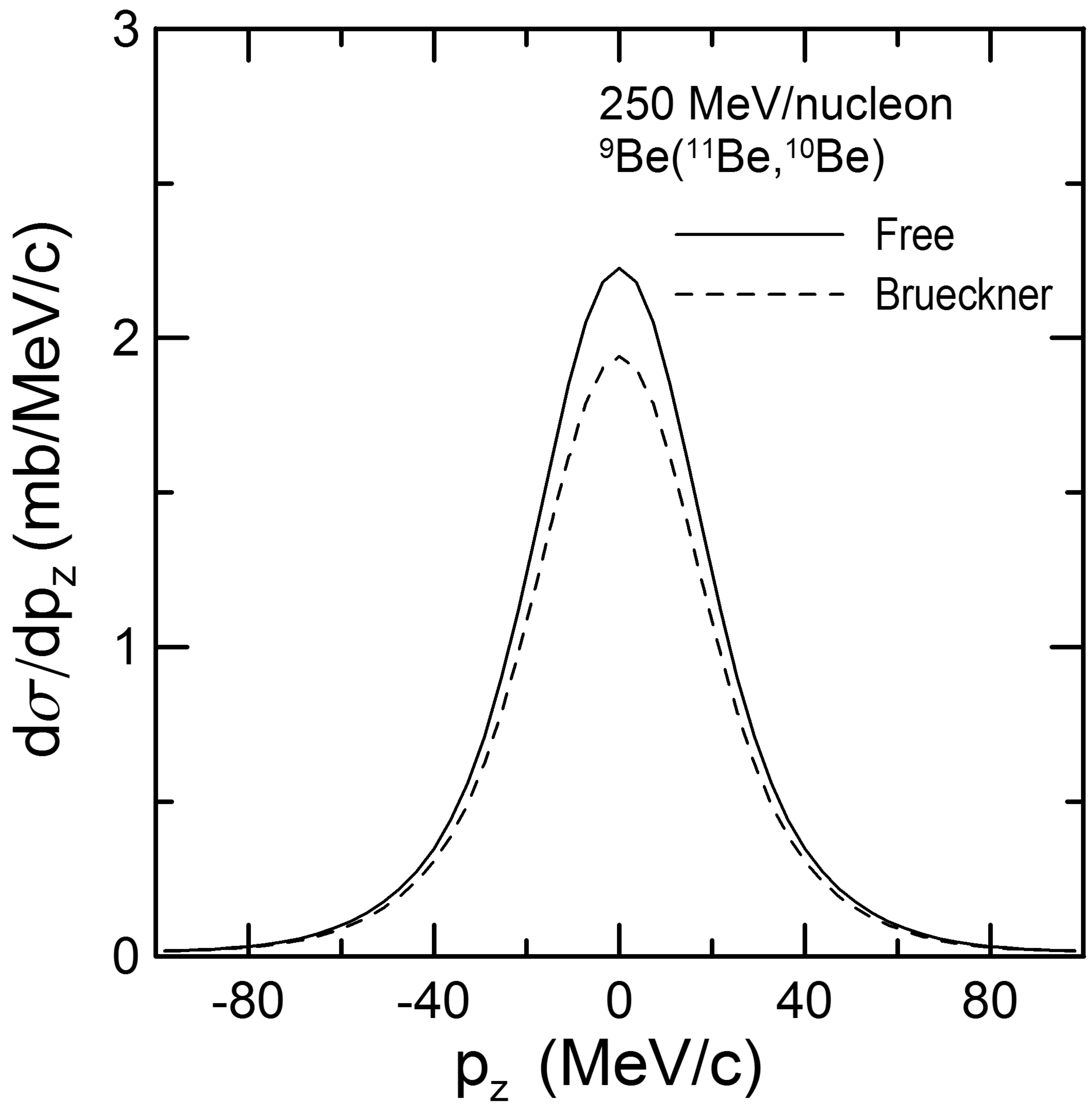}
\caption{Longitudinal momentum distribution for the residue in
the $^9$Be($^{11}$Be,$^{10}$Be),  reaction at 250 MeV/nucleon. The dashed curve is the cross
section calculated using the NN cross section from the Brueckner theory and the solid curve is obtained  the free cross section.}
\label{dsdpz}
\end{center}
\end{figure}

In figure \ref{dsdpz}  we plot the longitudinal momentum distributions for the reaction $^9$Be($^{11}$Be,$^{10}$Be),  at 250 MeV/nucleon \cite{BC10}. The dashed curve is the cross section calculated using the NN cross section from the Brueckner theory and the solid curve is obtained  the free cross section. One sees  that the momentum distributions are reduced by 10\%, about the same as  the total cross sections, but the shape remains basically unaltered. If one rescales the dashed curve to match the solid one, the differences in the width are not visible \cite{Kar11}. 

\section{Conclusions}
There were many questions not addressed in this review, such as the role of central nucleus-nucleus collisions in determining phase transition, equation of state, and a quark-gluon plasma, all topics or relevance in astrophysics. The review was more focused on the role of short-lived, exotic nuclei. The important scientific questions to be addressed both experimentally and theoretically in nuclear physics of exotic nuclei with relevance for astrophysics comprise: (a) How do loosely-bound systems survive and what are the general laws of their formation and destruction? (b) Are new types of radioactivity possible? (c) Are new types of nuclear symmetry and spatial arrangement possible? (d) What are the limits of nuclear existence? (e) How do the properties of nuclear matter change as a function of density, temperature and proton-to-neutron ratio? (f) How do thermal and quantum phase transitions occur in small systems? (g) What determines the shape and symmetry properties of an exotic nucleus? (h) How does quantum tunneling of composite particles occur in the process of reactions and decay? (i) What are the manifestations of fundamental forces and symmetries in unusual conditions? (j) How were the elements heavier than iron formed in stellar explosions? (k) How do rare isotopes shape stellar explosions? (l) What is the role of rare isotopes in neutron stars?
These questions provide extreme challenges for experiments and theory. On the experimental side, producing the beams of radioactive nuclei needed to address the scientific questions has been an enormous challenge. Pioneering experiments have established the techniques and present-generation facilities have produced first exciting science results, but the field is still at the beginning of an era of discovery and exploration that will be fully underway once the range of next- generation facilities becomes operational. The theoretical challenges relate to wide variations in nuclear composition and rearrangements of the bound and continuum structure, sometimes involving near-degeneracy of the bound and continuum states. The extraction of reliable information from experiments requires a solid understanding of the reaction process, in addition to the structure of the nucleus. In astrophysics, new observations, for example the expected onset of data on stellar abundances, will require rare-isotope science for their interpretation.

\medskip 
Supported by the US-DOE grants DE-FG02-08ER41533, DE-FG02-10ER41706, and DE- FC02- 07ER41457 (UNEDF, SciDAC-2) and the Research Corporation.

\end{document}